\documentclass[aps,prl,twocolumn,a4paper,10pt,notitlepage,footinbib,superscriptaddress,showpacs]{revtex4-1}%
\usepackage[english]{babel}
\usepackage[T1]{fontenc}
\usepackage{endnotes}
\usepackage{amssymb,amsmath,amsfonts}
\usepackage{textcomp}
\usepackage{graphicx,color}
\usepackage[utf8x]{inputenc}
\usepackage{float}
\usepackage{placeins}
\usepackage{multirow}
\usepackage{colortbl}
\usepackage{tabulary}
\usepackage{etoolbox}
\begin{document}

\title{Chaotic and periodical dynamics of active chiral droplets}

\author{Livio Nicola Carenza}
\affiliation{Dipartimento  di  Fisica,  Universit\`a  degli  Studi  di  Bari  and  INFN,  via  Amendola  173,  Bari,  I-70126,  Italy}

\author{Giuseppe Gonnella} 
\affiliation{Dipartimento  di  Fisica,  Universit\`a  degli  Studi  di  Bari  and  INFN,  via  Amendola  173,  Bari,  I-70126,  Italy}

\author{Davide Marenduzzo}
\affiliation{SUPA, School of Physics and Astronomy, University of Edinburgh, Mayfield Road, Edinburgh EH9 3JZ, United Kingdom}

\author{Giuseppe Negro}
\affiliation{SUPA, School of Physics and Astronomy, University of Edinburgh, Mayfield Road, Edinburgh EH9 3JZ, United Kingdom}

\begin{abstract}
The interplay between the chirality of many biological molecules and the energy injected at small length-scales as the result of biological processes is at the base of the life of the cells. With the aim of unveiling the connection between these two features, here we analyze by means of lattice Boltzmann simulations the behavior of an active droplet of cholesteric liquid crystal under the effect of intense active doping, within the framework of active gel theory. 
We find that a droplet of chiral liquid crystal, fueled by active force dipoles, develops defect loops (closed disclination lines) that pierce the interior of the droplet, leading the droplet to develop an erratic motility mode. When the droplet is fueled by in-warding active torque dipoles, three different dynamical regimes develops at varying both the thermodynamic chirality and the strength of active energy injection: a stable rotational state at low activity, an intermittent disclination dance regime, and a turbulent state where closed disclination lines formation is inhibited and new pairs of oppositely charged surface defects leads to the development of chaotic rotational motion. Finally, we show that out-warding torque dipoles are able to sustain a periodical dynamics at higher chirality characterized by the nucleation/annihilation of pairs of disclination rings.
\end{abstract}

\maketitle

\section{Introduction}
\label{intro}
Biological matter organizes in complex structures~\cite{ramaswamy2010,marc2013,Zhou1265} to guarantee the correct functioning of the processes at 
the base of life, even in the smallest living organisms, \emph{i.e.} prokaryotic cells.
For instance, DNA develops a double chiral structure that allows for strong and flexible morphologies 
which are important for the dynamics of DNA replication and transcription.
Some species of bacteria, such as \emph{E. Coli}, are equipped with filamentous proteins--\emph{flagella}--protruding from their body.
These often exhibit a helicoidal structure that, when twisted, acts in a fan-like fashion, thus allowing bacteria to move~\cite{Purcell1997,Riedel2005}.
Many other cytoskeletal proteins aggregate into chiral structures or perform twisting movements~\cite{Wang2012,naganathan2014}, as in the case of acto-myosin and microtubules, that are implemented by cells for structural stability and axonal transport~\cite{Sabry1995} (the mechanism responsible for the movement of organelles through the cytoplasm).
These biological systems mostly evolve in fluidic environments where hydrodynamic interactions play a fundamental role in the aforementioned processes
and the dynamics of chiral structures may result into a source of angular momentum on the surrounding fluid.
Therefore, understanding the relevance of chirality in such non equilibrium systems is crucial for capturing the physics of cellular life.

This particular kind of soft matter falls under the classification of \emph{active matter}~\cite{marc2013,carenza2019}, comprising those systems that are capable to  convert internal or chemical energy into motion. In the last decades this field gathered the attention of the physics community both for the possible application in the design of new smart materials, and for its connection to more fundamental problems related to the physics of non-equilibrium phenomena \cite{marchetti2013,doostmohammadi2018,negro2018,C9SM01288E,cates2008,loisy2018,giomi2010,hatwalne2004,saintillan2018,doi:10.1063/1.5090059}. 
For instance, the individual constituents of active gels (\emph{i.e.} bacteria, acto-myosin or microtubules bundles) exhibit the natural tendency to align and assembly in a nematic (or polar) fashion, thus giving rise to an \emph{active} liquid crystal.
Both experiments and numerical simulation on this kind of active systems have highlighted the existence of three dynamical states, which can be selected by tuning the rate at which energy is introduced in the system by the action of the active constituents. For small activity, when reactive and dissipative effects are capable of absorbing the energy injected, the system settles into a quiescent state, similar to its equilibrium counterpart where no significant flow can be detected. As activity is increased, active injection is capable of exciting bending deformations of the liquid crystal pattern that behave as energy sources. This regime is usually characterized by the occurrence of patterned flows and is typically addressed as \emph{spontaneous flow}~\cite{Edwards_2009,sanchez2012,simha2002,kruse2004,marenduzzo2007,PhysRevLett.101.198101}. By further increasing activity, the system enters a chaotic state, named \emph{active turbulence}~\cite{dombrowski2004,wensink2012,carenza19turb,Doostmohammadi2017,PhysRevLett.122.168002,PhysRevLett.110.228102,copar2019}, characterized by defect proliferation and time-dependent flows resembling the ones usually observed in turbulent fluids flowing at high Reynolds number. The origin of such phenomenon is drastically different from usual turbulence, where chaos emerges as the result of energy transfer between different length-scales due to advective mechanisms. On the contrary, active systems flow at negligible Reynolds numbers, where viscous effects dominate over advection~\cite{carenza19turb} and the resulting chaotic dynamics is due to the non-trivial interaction between the liquid crystal and the underlying fluid.
We will continue to refer to this regime as \emph{turbulent} or \emph{chaotic}, as customary in works on active matter.

Many models have been advanced to describe the behaviour of active matter, and in particular active fluids, ranging from 
particle-based approaches~\cite{bechinger2016,digregorio2018,doi:10.1142/S0129183114410046,cates2015,toner2005} to mesoscopic ones~\cite{PhysRevLett.115.188302,PhysRevX.8.031080,carenza2019,Negro_2019_epl,bonelli2019,carenza2019intj}. Among the latter, the active gel theory was found to successfully reproduce a 
certain number of behaviors found in experiments~\cite{negro2018,doostmohammadi2018,cates2018,PhysRevLett.112.147802,PhysRevX.9.041047,Giomi_2012}. Here, the orientational features of the active agents are encoded 
in a vector or tensor field (according to the symmetries of the particular system under consideration), whose dynamics is coupled to the hydrodynamics of the underlying fluid.
Activity is usually introduced by means of a coarse-grained description of the stress exerted by constituents on the surrounding environment~\cite{simha2002,marc2013}. 
%Simulations on different systems, confirmed that different dynamical states occur at changing the intensity of the active doping. For small activity, when reactive and dissipative effects are capable of absorbing the energy injected, the system settles into a quiescent state, similar to its equilibrium counterpart where no significant flow can be detected. As activity is increased, active injection is capable of exciting bending deformations of the liquid crystal pattern that behave as energy sources. This regime is usually characterized by the occurrence of patterned flows and is typically addressed as \emph{spontaneous flow}~\cite{Edwards_2009}. By further increasing activity, the system enters a chaotic state, named \emph{active turbulence}~\cite{Doostmohammadi2017,copar2019}, characterized by defect proliferation and time-dependent flows resembling the ones usually observed in turbulent fluids flowing at high Reynolds number. Nevertheless, the origin of such phenomenon is drastically different from usual turbulence, where chaos emerges as the result of energy transfer between different length-scales due to advective mechanisms. On the contrary, active systems flow at negligible Reynolds numbers, where viscous effects dominate over advection~\cite{carenza19turb} and the resulting chaotic dynamics is due to the mutual interaction between the liquid crystal and the underlying fluid.
 
Recently, we made use of the active gel theory to analyse the behavior of an active cholesteric liquid crystal confined in a droplet embedded in an isotropic passive fluid~\cite{Carenza22065}. In presence of tangential anchoring of
the liquid crystal at the interface, we found that the droplet develops different behaviors
at varying both chirality and the intensity of activity.
In particular, a passive nematic droplet (where  activity and chirality are both null) sets into a configuration that minimizes the free energy of the system (see later), characterized by two boojums ($+1$ aster-like point defects) at the poles.
By increasing activity over a certain threshold, the nematic pattern on the droplet equator develops bending deformations that induce a regular rotational motion around the axis defined by the two boojums.
A similar picture also applies for chiral droplets, but only at small cholesteric powers.
Indeed, if the chirality of the liquid crystal is strong enough, the mutual effect of elasticity and activity leads to the recombination of defects: starting  from the poles, the two boojums move close to each other with the inner structure of the liquid crystal characterized by a line of strong twist connecting the center of the droplet with the two surface defects--a configuration reminiscent of the well-known Frank-Pryce structure~\cite{zumer,FrankPrice}. Activity still powers the rotation of the droplet, which, due to its asymmetric pattern, is linearly propelled in the direction normal to the plane of rotation of the two boojums.

In our previous paper we mostly addressed this low/middle range of activity. Here, we will focus on the active turbulent state which can be reached both rising the activity (for any chiral power) and/or chirality (even at very small activity).
In the following we will discuss the effect of confinement and chirality on the dynamical state of the droplet. We will show how the momentum supplied by activity is used to generate disclination loops in the droplet interior and how their creation/annihilation dynamics fuels the erratic motion of the droplet itself. Moreover, we will consider the case of an active term in the stress tensor which acts as a source of angular momentum. In this case, we find that, for low activity, the relative handedness of thermodynamic and non-equilibrium chirality may trigger the onset of different non-equilibrium steady-states characterized by the intermittent dynamics of the active liquid crystal. By increasing the strength of active torque dipoles, the droplets enters a chaotic state, where the formation of disclination loops is inhibited and the droplet does not develop translational motion. Nevertheless, further pairs of oppositely charged defects develop on the surface, leading to chaotic rotational motion.

\section{Model and Numerical Methods}
\label{sec-1}
In order to address the features of an active liquid crystal droplet suspended in a Newtonian fluid we introduce a scalar field $\phi$ to describe the concentration 
of active material and the $Q$-tensor, whose principal eigenvector $\mathbf{n}$--namely the director field--defines the direction at which the active constituents point, on average, in a certain position in space. In the uni-axial approximation we can write
\begin{equation}
Q_{\alpha \beta} = \dfrac{S}{2}(3 n_\alpha n_\beta -\delta_{\alpha \beta}),
\label{eqn:uniax}
\end{equation}
where the scalar field $S$ accounts for the local degree of order of the liquid crystal.
The dynamics of both order parameters, $\phi$ and $Q_{\alpha \beta}$ is coupled to the dynamics of the underlying fluid whose velocity will be denoted with $\mathbf{v}$.
In the following Section we will present the free energy of the system and the dynamical equations which rule the evolution of 
the dynamical fields. Later on, we will summarize the main features of the hybrid lattice Boltzmann approach that we made use of to time-integrate the equations.

\subsection{The model}
\label{sec-2}

%We considered an incompressible fluid with mass density $\rho$ and divergence-free velocity $\mathbf{v}(\mathbf{r},t)$.
%, subject to the condition $\nabla \cdot \mathbf{v}=0$.
%To characterize the state of the system we introduced two order parameters: a scalar conserved concentration field $\phi(\mathbf{r},t)$ and the $\mathbf{Q}$-tensor that respectively account for the concentration of active material and its orientational order.
%Generally $\mathbf{Q}$ is a traceless symmetric tensor, but in the uniaxial approxiamtion $\mathbf{Q}= \mathbf{n} \mathbf{n} - \frac{ \mathbf{I}}{3}$, where $\mathbf{n}$ is the \emph{director}, a unit vector expressing the local average orientation of the liquid crystal moleculs.
The equilibrium properties of the system are described by the following Landau-De Gennes free energy functional~\cite{degennes1993}
\begin{multline}
\mathcal{F}\left[\phi,Q_{\alpha \beta}\right] = \int dV \ \left[ \dfrac{a}{4} \phi^2 (\phi-\phi_0)^2 + \dfrac{k_\phi}{2} (\nabla \phi)^2
 \right. \\ \left.
+ A_0 \left[ \dfrac{1}{2} \left(1 - \dfrac{\chi(\phi)}{3} \right)\mathbf{Q}^2 -  \dfrac{\chi(\phi)}{3} \mathbf{Q}^3 +  \dfrac{\chi(\phi)}{4} \mathbf{Q}^4 \right] 
 \right. \\ \left.
+ \dfrac{K_Q}{2} \left[ (\nabla \cdot \mathbf{Q})^2 + (\nabla \times \mathbf{Q} + 2 q_0 \mathbf{Q})^2 \right]
+W (\nabla \phi) \cdot \mathbf{Q} \cdot (\nabla \phi) 
 \right].
 \label{eqn:freeE}
\end{multline}
The first term in the free energy describes a double-well potential whose minima are found in $0$ and $\phi_0$, which correspond to regions poor and rich of active material, respectively. Here the parameters $a$, $k_\phi$ control the surface tension and the interface width of the droplet, which ensure the droplet to maintain an approximatively spherical shape, despite active injection. The second line describes the first-order isotropic-nematic phase transition~\cite{degennes1993} so that the liquid crystal (LC) is confined in those regions where $\chi(\phi)=\chi_0 + \chi_s \phi > 2.7$, with $\chi_0= 10 \chi_s = 2.5$. The energy cost of elastic deformations of the LC pattern is modelled by the terms proportional to $K_Q$, where we adopted the single elastic constant approximation.
Chirality is introduced at an equilibrium level by adding a term proportional to the curl of the $Q$-tensor that favours the twisting of the LC. The parameter $q_0$ sets the strength of the chirality so that $|q_0|=2\pi/p_0$, where $p_0$ is the pitch of the cholesteric helix.
Right-handed chirality, as the one here considered, is achieved by requiring $q_0$ to be positive.
To compare the cholesteric features of the LC and the geometry in which is confined, we introduce the adimensional number $N=2R/p_O$ that counts the number of windings the LC director performs inside a droplet of radius $R$.
The last term in Eq.~\eqref{eqn:freeE} is an anchoring term that favours tangential anchoring when $W>0$.

The dynamical equation for the conserved concentration field is a convection-diffusion equation:
\begin{equation}
\partial_t + \nabla \cdot (\phi \mathbf{v}) = \nabla \cdot \left( M \nabla \dfrac{\delta \mathcal{F}}{\delta \phi} \right),
\label{eqn:conv-diff}
\end{equation}
where $M$ is the mobility parameter. The Beris-Edwards equation rules the dynamics of the Q-tensor:
\begin{equation}
(\partial_t + \mathbf{v}\cdot \nabla) \mathbf{Q} - \mathbf{S}(\mathbf{W},\mathbf{Q}) = \Gamma \mathbf{H},
\label{eqn:BerisEdwards}
\end{equation}
where $\mathbf{W}= \nabla \mathbf{v}$ and the strain-rotational derivative is given by
\begin{multline}
\mathbf{S}(\mathbf{W},\mathbf{Q}) = (\xi \mathbf{D} + \mathbf{\Omega})(\mathbf{Q}+\mathbf{I}/3) \\ + (\mathbf{Q}+\mathbf{I}/3)(\xi \mathbf{D} - \mathbf{\Omega}) - 2 \xi (\mathbf{Q}+\mathbf{I}/3) Tr (\mathbf{Q}\mathbf{W}),
\label{eqn:material_derivative}
\end{multline}
with $\mathbf{D}$ and $\mathbf{\Omega}$ respectively denoting the symmetric and asymmetric part of $\mathbf{W}$. The adimensional parameter $\xi$ controls the aspect-ratio of the liquid crystal molecules. In this paper we consider $\xi=0.7$ which corresponds to flow-aligning rod-like molecules. $\Gamma$ is the rotational viscosity, while the molecular field, driving the LC towards equilibrium is given by
\begin{equation}
\mathbf{H} = - \dfrac{\delta \mathcal{F}}{\delta \mathbf{Q}} + \dfrac{\mathbf{I}}{3} Tr \left( \dfrac{\delta \mathcal{F}}{\delta \mathbf{Q}} \right)
\end{equation}
Finally, the Navier-Stokes governs the flow evolution:
\begin{equation}
(\partial_t + \mathbf{v} \cdot \nabla ) \mathbf{v} = \nabla \cdot \left[ \mathbf{\sigma}^{pass} + \mathbf{\sigma}^{act} \right].
\label{eqn:Navier_Stokes}
\end{equation}
Here, we split the stress tensor contribution into a passive and an active term. The first one accounts for the dissipative and reactive contributions (stemming from the free energy and invariant under time reversal).
In more detail, we consider an isotropic pressure term
$$
\sigma_{\alpha \beta}^{hydro}= -p \delta_{\alpha \beta},
$$ 
with $p$ the hydrodynamic pressure.
Viscosity dissipation is given by:
$$
\sigma^{visc}_{\alpha \beta} = 2 \eta D_{\alpha \beta},
$$
with $\eta$ the shear viscosity.
The relaxation dynamics of the two order parameters affects the hydrodynamics through the following passive terms:
\begin{equation}
\sigma^{bm}_{\alpha \beta} = \left(f-\dfrac{\delta\mathcal{F}}{\delta \phi} \right)\delta_{\alpha \beta} - \dfrac{\delta \mathcal{F}}{\delta (\partial_\beta \phi)} \partial_\alpha \phi,
\end{equation}
where $f$ is the free energy density, and
\begin{multline}
\sigma_{\alpha \beta}^{el} = -\xi H_{\alpha \gamma} \left(Q_{\gamma \beta} + \dfrac{1}{3} \delta_{\gamma \beta} \right) -\xi  \left(Q_{\alpha \gamma} + \dfrac{1}{3} \delta_{\alpha \gamma} \right) H_{\gamma \beta} 	\\ + 2\xi \left(Q_{\alpha \beta} - \dfrac{1}{3} \delta_{\alpha \beta} \right) Q_{\gamma \mu} H_{\gamma \mu} + Q_{\alpha \gamma} H_{\gamma \beta}  - H_{ \alpha \gamma} Q_{\gamma \beta} .
\end{multline}
To introduce activity in the model, we consider a phenomenological active stress tensor~\cite{ramaswamy2010} given by:
\begin{equation}
\sigma_{\alpha \beta}^{act} = -\zeta \phi Q_{\alpha \beta} - \bar{\zeta} \epsilon_{\alpha \mu \nu} \partial_\mu (\phi Q_{\nu \beta}).
\end{equation}
Here, the first term is derived by coarse-graining the force exerted by the swimmers, proportional to the \emph{force dipole active parameter} $\zeta$. In the following we will limit our analysis to the case $\zeta>0$ corresponding to extensile entities that produce outward-pointing force dipoles on the fluid.
The second term, proportional to $\bar{\zeta}$ describes the source of angular momentum provided by torque dipoles that swimmers deploy on the surrounding environment. For positive $\bar{\zeta}$ the active torque corresponds to a pair of outwarding torques (analogue to the one which opens a bottle cap). In turn, $\bar{\zeta}<0$ describes an inward pair of torques (similar to the one used to close a bottle cap). Importantly, the non-equilibrium twist introduced by the activity may reinforce or oppose the handedness of the
thermodynamic twist, which is determined by the sign of $q_0$. Therefore, what is physically meaningful is the relative orientation of the thermodynamic handedness with respect to the one introduced by active torques.
For instance, in our paper we consider a right-handed LC by setting $q_0 > 0$. In this case, outwarding active torques ($\bar{\zeta}>0$) result into flows which strengthen the thermodynamic twist. Conversely, a negative active torque would have the opposite effect, countering the handedness of the LC. The opposite would happen in case of a left-handed cholesterics.

%A negative value of $\bar{\zeta}$ corresponds to an inward pair of torques similar to that used to open a bottle cap.

\begin{table}[t]
\centering
\begin{tabular}{@{}|l|l|l|@{}}
\hline
\multicolumn{1}{|c|}{Model parameters }   &
\multicolumn{1}{|c|}{Simulation units}   &
\multicolumn{1}{|c|}{Physical units}   \\\hline
Shear viscosity, $\eta$           & $5/3$  & $1.67 \ \text{kPas}$       \\
Elastic constant, $K_Q$           & $0.01$  & $50 \ \text{nN}$      \\
Shape factor, $\xi$               & $0.7$  & dimensionless       \\
Diffusion constant, $D=Ma$        & $0.007$  & $0.06\ \text{$\mu$} \text{m}^2 \text{s}^{-1}$     \\
Activity, $\zeta$    & $0-0.01$  & $(0-100) \ \text{kPa}$       \\ 
Active torque, $\bar{\zeta}$    & $0-0.05$  & $(0-50) \ \text{$\mu$N/m}$       \\ 
\hline
\end{tabular}
\caption{Mapping between physical and simulation units. Length-scale $l^* = 1 \mu m$, time-scale $t^* = 10ms$ and
force-scale $f^* = 1 nN$ are set to $1$ in LB units.}\label{tabel_units}
\end{table}

Despite an actual realization of the system that we consider here has not yet been realized, it is possible to map simulation parameters into physical units, on the base of similar experimental systems. This mapping can be obtained by putting in correspondence the grid spacing with the typical length-scale $l^*$ of the constituents ($\sim 1 \mu m$ for microtubules suspensions); the time scale $t^*= 10 ms$ is chosen so to have a reference speed $v^*=l^*/t^* \sim 100 \mu m /s$ compatible with the typical intensity of flows in active cytoskeletal (or bacterial) suspensions, while the force scale $f^*=1nN$ is set to match the typical stress exerted by active constituents on the surrounding fluid. All other quantities can be obtained by rescaling the units of measure with respect to $l^*, t^*, f^*$. Some relevant parameters are reported in Table~\ref{tabel_units}.

\subsection{The Numerical Method}
\label{sec-3}
The dynamical equations have been integrated by means of a  hybrid lattice Boltzmann (LB) method~\cite{succi2001} which combines a predictor-corrector LB treatment for the Navier-Stokes equation~\cite{denniston2001,carenza2019}  with a finite-difference algorithm to solve the order parameters dynamics, implementing a first-order upwind scheme for the convection term, and fourth-order accurate stencil for the computation of space derivatives.

The evolution of the fluid is described in terms of a set of distribution functions ${f_i(\mathbf{r}_\alpha,t)}$ (where the index $i$ labels different 	ranging from $1$ to $15$) defined on each lattice site $\mathbf{r}_\alpha$. Their evolution follows a discretized version of the Boltzmann equation in the BGK approximation~\cite{doi:10.1142/S0129183119410055}:
\begin{multline}
f_i (\mathbf{r}_\alpha + \vec{\xi}_i \Delta t) - f_i (\mathbf{r}_\alpha,t) = \\ - \dfrac{\Delta t }{2} \left[ \mathcal{C}(f_i,\mathbf{r}_\alpha, t) + \mathcal{C}(f_i^*,\mathbf{r}_\alpha+ \vec{\xi}_i \Delta t, t) \right].
\label{eqn:LBevolution}
\end{multline}
Here $\lbrace \vec{\xi}_i \rbrace$ is the set of lattice velocities, that for the $D3Q15$ scheme implemented here are $\vec{\xi}_0=(0,0,0)$, $\vec{\xi}_{1,2}=(\pm u,0,0)$, $\vec{\xi}_{3,4}=(0,\pm u,0)$, $\vec{\xi}_{5,6}=(0,0,\pm u)$, $\vec{\xi}_{7-15}=(\pm u, \pm u, \pm u)$, where $u$ is the lattice speed. The distribution functions $f^*$ are first-order estimations to  $f_i (\mathbf{r}_\alpha + \vec{\xi}_i \Delta t) $ obtained by setting $f_i^* \equiv f_i$ in Eq.~\eqref{eqn:LBevolution}, and $\mathcal{C}(f,\mathbf{r}_\alpha, t)=-(f_i-f_i^{eq})/\tau + F_i$ is the collisional operator in the BGK approximation that can be expressed in terms of the equilibrium distribution functions $f_i^{eq}$ and supplemented with an extra forcing term for the treatment of the anti-symmetric part of the stress tensor. The physical hydrodynamical observables, \emph{i.e.} density and momentum of the fluid, are defined in terms of the distribution functions as follows:
\begin{equation}
\sum_i f_i = \rho \qquad \sum_i f_i \vec{\xi}_i = \rho \mathbf{v}.
\label{eqn:variables_hydro}
\end{equation}
Analogous relations also hold for the equilibrium distribution functions, thus ensuring mass and momentum conservation. To correctly reproduce the Navier-Stokes equation we impose the following condition on the second moment of the equilibrium distribution functions:
\begin{equation}
\sum_i f_i \vec{\xi}_i \otimes \vec{\xi}_i = \rho \mathbf{v} \otimes \mathbf{v} -\tilde{\sigma}^{bm} - \tilde{\sigma}^{el}_s,
\label{eqn:constrain_second_moment}
\end{equation}
and on the force term:
\begin{align}
&\sum_i F_i = 0, \nonumber \\ &\sum_i F_i \vec{\xi}_i = \mathbf{\nabla} \cdot (\tilde{\sigma}^{el}_a + \tilde{\sigma}^{act}), \\  &\sum_i F_i \vec{\xi}_i \otimes \vec{\xi}_i = 0, \nonumber
\label{eqn:constraint_force}
\end{align}
where we denoted with $\tilde{\sigma}^{el}_s$ and $\tilde{\sigma}^{el}_a$ the symmetric and anti-symmetric part of the polar stress tensor, respectively. The equilibrium distribution functions are expanded up to the second order in the velocities:
\begin{equation}
f_i^{eq} = A_i + B_i (\vec{\xi} \cdot \mathbf{v}) + C_i |\mathbf{v} |^2 + D_i (\vec{\xi} \cdot \mathbf{v})^2 + \tilde{G}_i : (\vec{\xi} \otimes \vec{\xi}).
\end{equation}
Here coefficients $A_i, B_i,C_i,D_i,\tilde{G}_i$ are to be determined imposing conditions in Eqs.~\eqref{eqn:variables_hydro} and \eqref{eqn:constrain_second_moment}. In the continuum limit the Navier-Stokes equation is restored if $\eta=\tau/3$.

We made use of a parallel approach implementing Message Passing Interface (MPI) to parallelize the code. We divided the computational domains in slices, and assigned each of them to a particular task in the MPI communicator. Non-local operations (such as derivatives), have been treated through the \emph{ghost-cell} approach~\cite{MPI}.
Simulations have been performed on a cubic lattice of size $L=192$.
%We initialized the system with a droplet of concentration field ($\phi=\phi_0$). 
The $Q$-tensor was randomly initialized inside the droplet and set to $0$ outside.
The values of free energy parameters are $a=0.07$, $k_{\phi}=0.14$, $A_0=1$, $K_Q=0.01$, and $W=0.02$. The rotational diffusion constant $\Gamma$ is set to  
$2.5$, while the diffusion constant to $M=0.1$.
All physical observables have been written in lattice units, as usual in computational works on active matter.

We initialize a droplet of concentration field with radius $R=32$, by setting $\phi=\phi_0$ inside the sphere and $0$ outside. The liquid crystal is set to zero outside the droplet, while inside is initialized with the uniaxial profile of Eq.~\eqref{eqn:uniax}, with $S=1$ and the director field $\mathbf{n}$ randomly oriented. The radius of the droplet has been chosen so to allow the cholesteric helicoidal pattern to smoothly vary in the droplet bulk and ensure numerical accuracy in derivatives computation.

The angular velocity of the droplet has been computed as:
$ \vec{\omega} = \int \text{d}\mathbf{r} \phi \dfrac{\Delta \mathbf{r} \times \Delta \mathbf{v}}{|\Delta \mathbf{r}|^2},$ where $\Delta \mathbf{r}= \mathbf{r} - \mathbf{R}$ and $\Delta \mathbf{v}= \mathbf{v} - \mathbf{V}$, with $\mathbf{R}$ and $\mathbf{V}$ respectively the position and the velocity of the center of mass of the droplet.
The identification of defects in the LC structure was carried out implementing the Westin metrics~\cite{Jankun-Kelly2009}.
%Don't forget to give each section, subsection, subsubsection, and
%paragraph a unique label (see Sect.~\ref{sec-1}).

\section{Active turbulence and disclination line proliferation in  cholesteric droplets}
\label{sec-4}

Much effort has been recently spent to understand the motility properties of active nematic (achiral) droplets~\cite{joanny2012,PhysRevLett.112.147802,activenematicdroplet,doostmohammadi2018,C9SM01076A} and polar droplets ~\cite{Tjhung2017,Whitfield2016,fialho2017,tjhung2012}.
Many factors are capable of influencing the dynamical state of active droplets, ranging from surface tension, anchoring strength and kind of active forcing, even if most of them have been analyzed exclusively for achiral liquid crystals in bidimensional environments. In particular, a bidimensional nematic droplet ($N=0$) is able to develop spontaneous motion for extensile active forcing either in the absence of anchoring or when the LC is tangentially anchored at the interface. This is because, active injection favors bending deformations of the extensile liquid crystal, thus leading to an asymmetric internal configuration resulting in a net source of momentum. On the contrary, homeotropic anchoring triggers the formation of local distortions of the droplet leading to rotation. In contractile systems, the liquid crystal is unstable under splay deformations and the interplay between anchoring and activity is found inverted, so that tangential anchoring favors rotation and hometropic or no anchoring lead to translational motion. 
Nevertheless, bidimensional droplets can also develop erratic motion if asymmetries arise as the result of strong active forcing. In this case the motion becomes erratic and no stable pattern can be recognized in the LC arrangement.
We will now present the results of our study, concerning the behavior of an active cholesteric droplet with tangential anchoring evolving in a fully 3D environment.
We will mostly address the high activity limit, where the amount of energy injected at small scales drives the development of chaotic flows.
In the next Section we will treat the case of a cholesteric droplet fueled by 
active force dipoles, while the further following Sections will be devoted to the effect of active torque dipoles.

\subsection{Active force dipoles}
\label{sec-5}

\begin{figure*}[t]
\centering
{\includegraphics[width=1\textwidth]{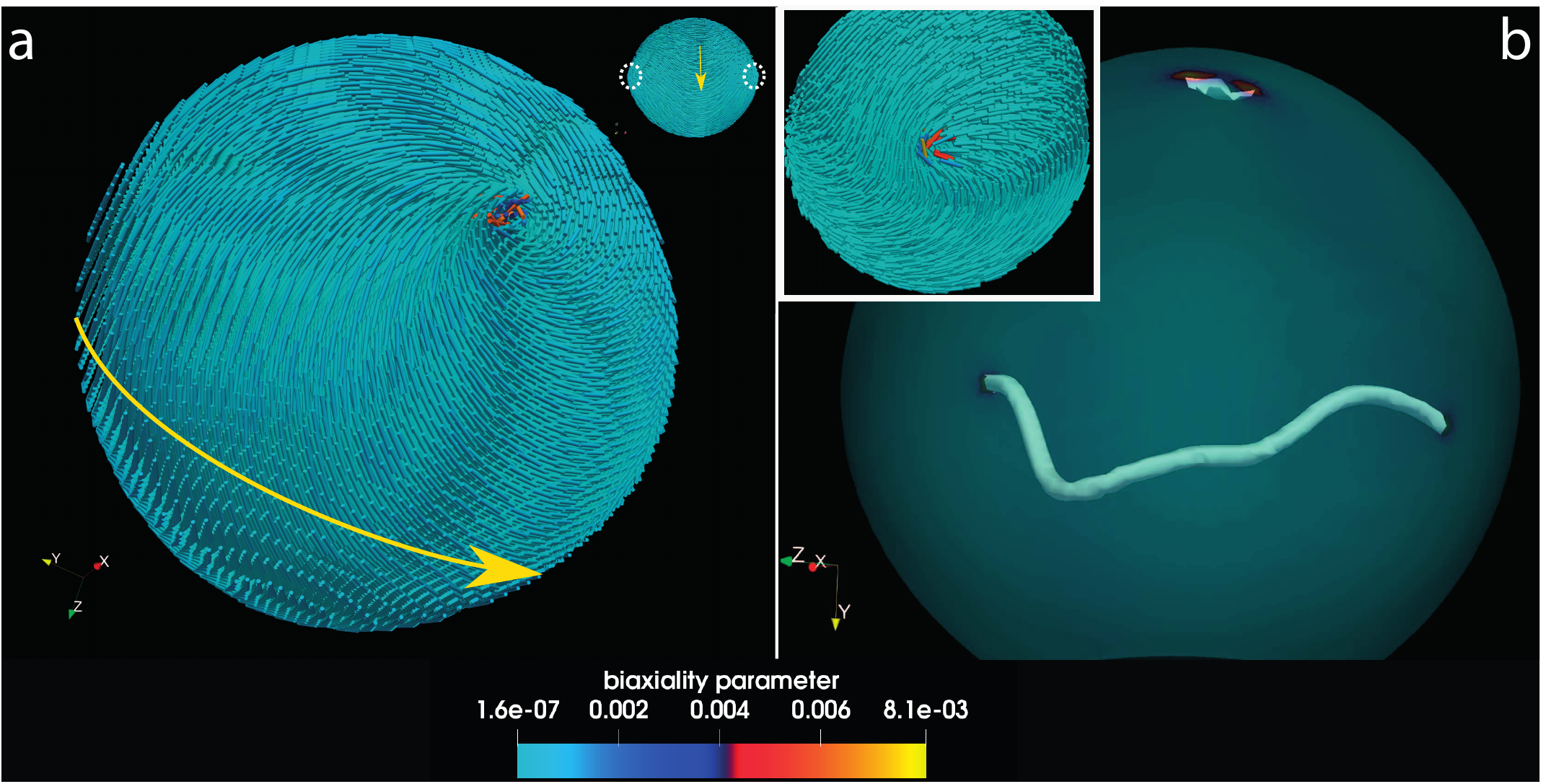}}\caption{\textbf{Spontaneous flow and active turbulence in a nematic droplet:} Panel (a) shows the nematic director on the droplet surface undergoing bending instability for $\zeta>3\times 10^{-4}$.  The deformation is greater at the droplet equator and acts as a source of momentum that fuels the regular rotation of the droplet around the axis defined by the two boojums at the droplet poles. Panel (b) shows a typical configuration of defect lines in the turbulent regime (here $\zeta=1.4\times10^{-3}$).  A disclination of semi-integer charge joins a pair $+1/2$ point defects on the surface (see inset) piercing the interior of the droplet. Fueled by activity, two defects may eventually overcome elastic repulsion and (momentarily) join to form a $+1$ surface defect (as in the top part of the droplet), resulting in the vanishing of the associated disclination line. For stronger active doping further disclination loops may be produced in the bulk.}
\label{fig-1}       % Give a unique label
\end{figure*}

Before getting involved in the discussion of the results concerning the behavior
of a cholesteric droplet under strong active forcing, we will consider the case at $N=0$, namely the nematic case.
In absence of activity (not shown here), the droplet relaxes into a configuration that minimizes the free energy in Eq.~\eqref{eqn:freeE}, characterized by the presence of two antipodal boojums, in accordance with the Gauss-Bonnet theorem stating that the total topological charge of a vector field tangentially anchored on a sphere is constrained to be $+2$.
In our previous paper~\cite{Carenza22065} we announced that, as the activity parameter $\zeta$ is progressively increased, the droplet undergoes different dynamical regimes: first a quiescent regime, where no difference can be effectively detected from the passive case, a rotational regime (see panel (a) in Fig.~\ref{fig-1}), where the droplet sets up a stationary rotational motion around the axis connecting the two boojums and, finally, a turbulent regime.

For an active LC confined in a droplet, the most relevant control parameter is the dimensionless activity $\theta = \zeta R^2 /K$, which measures the importance of the rate of energy input with respect to elastic relaxation. Indeed, the mechanism at the base of the self-sustained flows developing in the system is the instability of extensile active nematics to bending deformations. 
Regions of strong deformations, such as point or line defects and/or bendings of the LC, act as a source of momentum, being the active force density proportional to the gradients of the $Q$-tensor ($f\propto \nabla \cdot \mathbf{Q}$).
As long as elastic and dissipative phenomena can absorb and dissipate the energy injected, a non-equilibrium stationary regime is reached and eventually the droplet sets up into the stable rotational motion previously mentioned.
In such regime the dynamics is basically driven by the mutual balancing of active injection and viscous dissipation, since elastic absorption is highly suppressed, having the LC attained a stationary pattern.
But what if the rate at which energy is injected is not compensated by dissipation?
In this case the excess energy may strengthen elastic deformations, which result into stronger active forces, hence in the increment of the total  kinetic energy.  Active flows are then capable to advect the LC, pulling it apart from its stable orbit, resulting in further deformations leading to further momentum source. The consequent dynamics is no more periodic and the flows become chaotic.

The deformations of the nematic pattern on the surface leads to the splitting of each of the two antipodal boojums into two pairs of $+1/2$ point defects which are connected by a disclination line that pierces the interior of the droplet (see Fig.~\ref{fig-1}(b)).
The energy deployed due to the strong deformation in the neighbourhood of the defect lines favours the formation of new defects even in the droplet interior.
Despite both point and line defects are topologically stable in a full $3D$ geometry, we found that closed disclination lines are more likely to form for energetic reasons. 
These disclination loops undergo a chaotic evolution and eventually merge or annihilate with other loops evolving in the bulk. 
This phenomenology has important effects on the overall dynamics of the droplet.
Indeed, the asymmetric configuration of LC patterns leads to a net injection of momentum which results into the rambling motion of the droplet, depending on the inner defect structure.
\begin{figure*}[t]
\centering
{\includegraphics[width=0.9\textwidth]{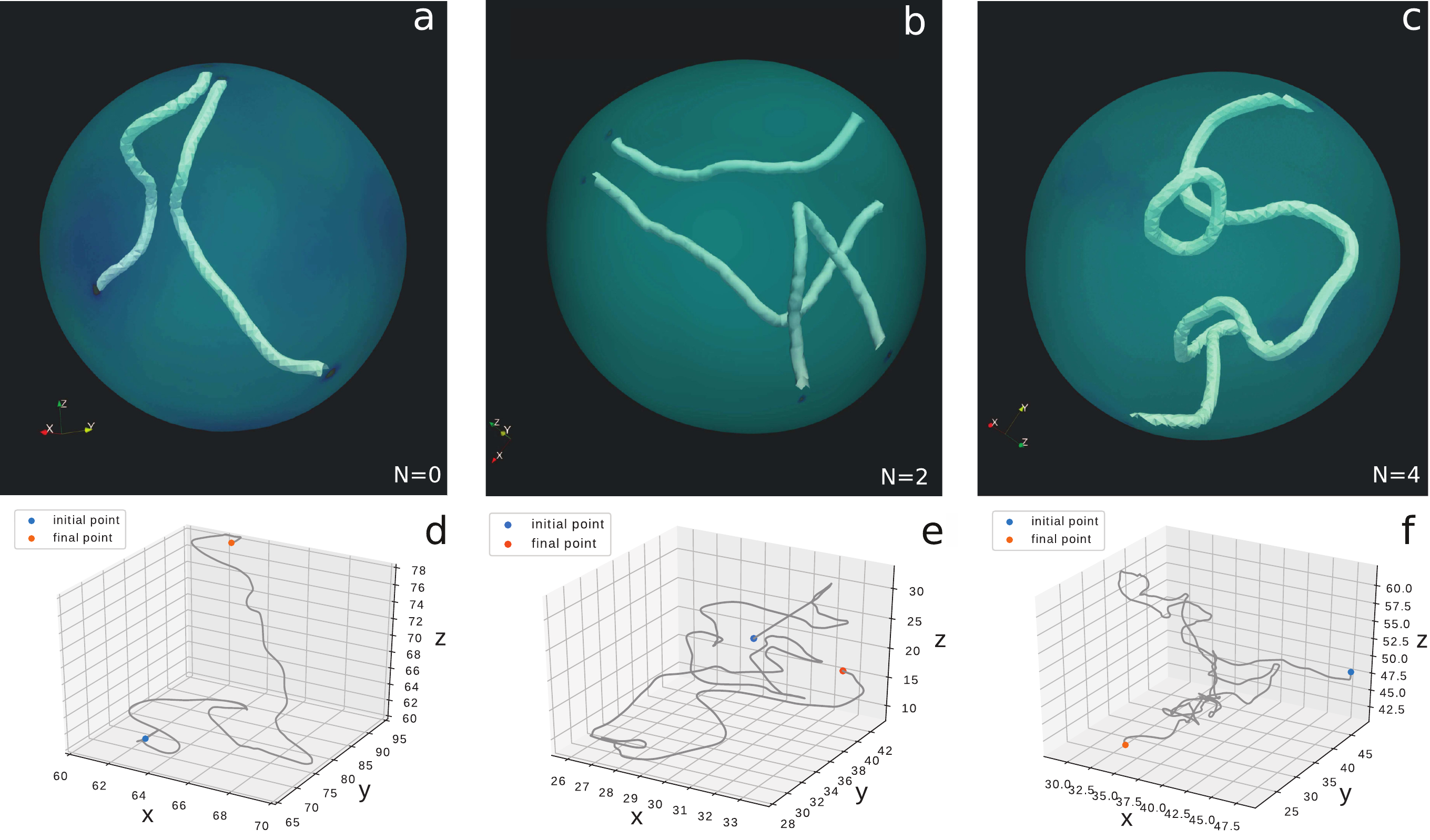}}\caption{\textbf{Erratic motility of cholesteric droplets in the active turbulent regime}: Top panels (a-c) show the contour plot of droplets and their defect lines for different cholesteric strengths ($N=0,2,4$ respectively) at the same value of the activity parameter $\zeta=2\times 10^{-4}$. The structure of the disclination lines becomes progressively more and more curved as chirality is increased. This has important effects on the motility properties of the droplet. Bottom panels (d-f) show the trajectory of the centers of mass of the droplets shown in the top panels in an observation range of $1.2\times 10^{6}$ LB iterations. As chirality is increased the motion becomes more chaotic and changes in direction more and more frequent.}
\label{fig-2}       % Give a unique label
\end{figure*}

We now take into account the effect of active force dipoles on a \emph{cholesteric} droplet. Even in this case we found three different regimes with some important differences with respect to the nematic case.
%At negligible activity, the liquid crystal arranges in a twsting fashion to minimize the free energy.
The LC bulk configuration is characterized by two antipodal boojums as previously described for a nematic droplet. Nevertheless, in this case twisting deformations develop even in absence of activity, to minimize the free energy. 
As activity is increased the droplet starts to rotate around the axis defined by the two boojums, but only for $N \lesssim 1.5$. At higher chirality, activity favors the recombination of the two defects that are pulled together as the effect of an activity-induced attraction. The steady-state configuration resembles the Frank-Pryce structure, usually found in cholesteric droplets at strong chirality ($N \gtrsim 5$). Though, the rotational state is preserved and while the two defects orbit around each other, the droplet is linearly propelled due to the asymmetry of the LC configuration in the interior of the droplet~\cite{Carenza22065}.

How does chirality influence the onset of the turbulent dynamics?
The mechanism is basically the same as in the nematic case previously analyzed: the energy that is injected due to active mechanisms favors the formation of new disclinations in the bulk (see Fig.~\ref{fig-2}(b-c)), favored by the distortions of the LC induced by the twisted cholesteric structure.
Consequently, as $N$ is increased, the structure of the defect lines becomes more and more distorted and defect formation/annihilation rate grows. As a result, the onset of the turbulent regime occurs at progressively lower values of the activity parameter $\zeta$.
The asymmetry thus generated causes the droplet to move due to unbalanced momentum injection. Moreover, chirality has an important effect on the trajectory followed by the droplet (see bottom panels in Fig.~\ref{fig-2}). Indeed, while in the nematic case the trajectory exhibits long lapses of linear motion, as chirality is increased, more and more changes in direction occur, in accordance with the observation that the dynamics of the defect lines in the droplet's bulk becomes correspondingly more chaotic.

\section{Active torque dipoles}
\label{sec-6}
We now discuss the case of a cholesteric droplet fueled by active torques only.
Analogously to the previous case, the behavior is determined by comparing the intensity of dipolar torque activity proportional to $\bar{\zeta}$ and the equilibrium properties of the LC, hence the adimensional number which best assesses the response of the system is given by $\bar{\theta} = \bar{\zeta} R /K$. 
The dynamics of our cholesteric droplet strongly depends on the sign of $\bar{\zeta}$. Indeed, while out-warding torque dipoles ($\bar{\zeta} >0$) strengthen the handedness of our (right-handed) LC, in-warding torque dipoles ($\bar{\zeta} <0$) counter it. Next Section focuses on the latter case, while the former is addressed in Section~\ref{sec-6.2}.

\subsection{In-warding active torque dipoles}
\label{sec-6.1}
\begin{figure*}[t]
\centering
{\includegraphics[width=1.\textwidth]{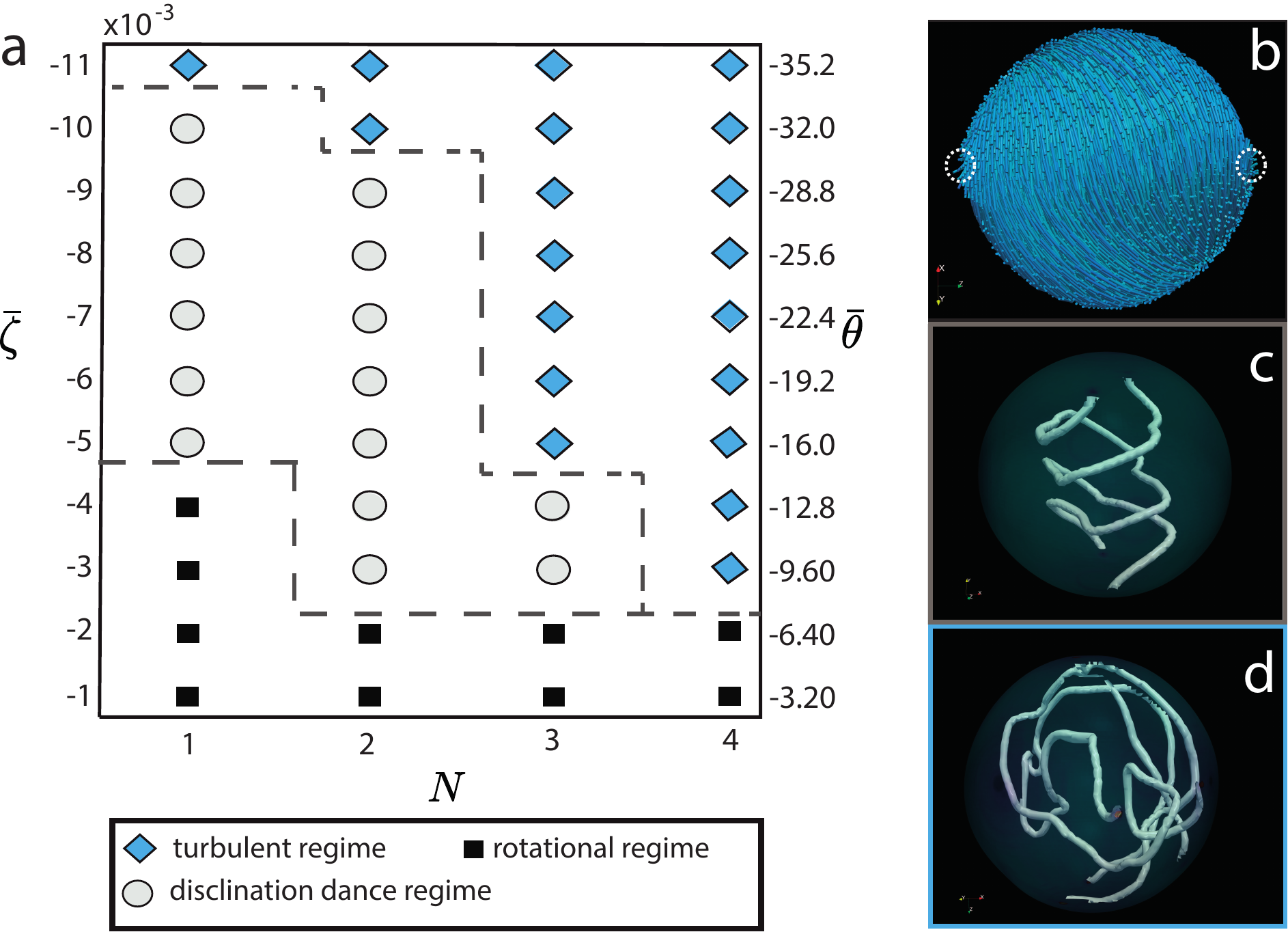}}
\caption{\textbf{Cholesteric droplets fueled by active torques only}: Panel (a) shows the different behaviors of the droplet found when varying both the intensity of in-warding active torques and the cholesteric power. Panel (b) shows the nematic configuration on the surface of the droplet in the rotational regime at $N=3$ and $\bar{\zeta}=-10^{-3}$. Panel (c) and (d) show the structure of the disclination lines connecting point defects on the droplet surface at $N=3$ respectively for $\bar{\zeta}=-3\times 10^{-3}$ (disclination dance regime) and $\bar{\zeta}=-8\times 10^{-3}$ (turbulent regime).}
\label{fig-3}       % Give a unique label
\end{figure*}

\begin{figure*}
\centering
{\includegraphics[width=1\textwidth]{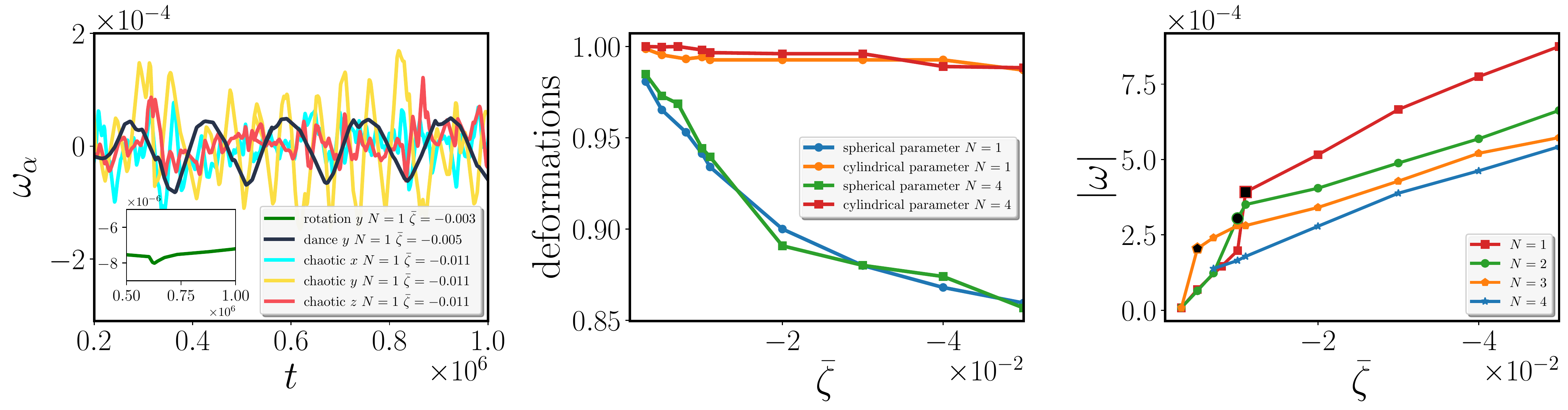}}
\caption{\textbf{Rotational velocity and deformations in droplets fueled by in-warding active torques}: Panel~(a) shows the typical profiles of the components of the angular velocity in the three possible dynamical regimes at varying the intensity of the in-warding active torque at $N=1$. The constant angular velocity for the rotational regime at $\bar{\zeta}=-3\times 10^{-3}$ is shown in the inset. The dark blue line shows the oscillating behavior of $\omega_n$ (being $\mathbf{n}$ the axis of rotation) in the disclination dance regime at $\bar{\zeta}= -5 \times 10^{-3}$, while other profiles refers to the Cartesian components of $\mathbf{\omega}$ in the chaotic regime at $\bar{\zeta}= -1.1 \times 10^{-2}$.  Panel~(b) shows the spherical and cylindrical parameters. Panel~(c) shows the time-averaged modules of the angular velocity at varying the intensity of the in-warding torque dipoles for different values of $N$. Black dots correspond to the first case in the chaotic region. }
\label{fig-4}       % Give a unique label
\end{figure*}

We start our discussion from the case of a cholesteric droplet fueled by in-warding active torque dipoles, where the competing mechanism between thermodynamic chirality and non-equilibrium twist gives rise to different dynamical regimes. For small values of the active torque parameter $|\bar{\zeta}|$, and due to the imposed tangential anchoring, the droplet sets up into a configuration characterized by the presence of two boojums at the poles, as shown in Fig.~\ref{fig-3}(b), regardless of the intensity of thermodynamic chirality, which induces a twisted bipolar pattern in the LC arrangement on the droplet surface. This configuration is roughly stationary and the droplet rotates around the axis defined by the two $+1$ surface defects, with approximately constant angular velocity (see inset of panel~(a) in Fig.~\ref{fig-4}).

By increasing $|\bar{\zeta}|$, active torques favour the splitting of each of the two $+1$ boojums into two $+1/2$ defects. 
This has important effects on the arrangement of the liquid crystal in the droplet bulk, since semi-integer defects on the surface are connected in pair  by a disclination line which pierces the droplet interior (see for instance Fig.~\ref{fig-3}(c)). The development of disclination lines has topological origin. Indeed, while line disclinations with integer topological charge are intrinsically unstable~\cite{chaikin2000principles},  disclinations with semi-integer charge are perfectly allowed and may generate from the splitting of a $+1$ point defect into two $+1/2$ defects.
The dynamical response of the droplet at moderate $|\bar{\zeta}|$ strongly depends on the intensity of thermodynamic chirality. In particular, for $N \leq 3$, we found that the disclination lines attain a regular spiral shape (Fig.~\ref{fig-3}(c)) in the grey region of the phase diagram in Fig.~\ref{fig-3}(a) and set up a periodical motion triggered by the mirror rotation of the two pairs of point defects. As the disclination lines wire around each other, they end up intersecting and recombining, thus leading to the mutual swap of the linked defect pairs.  This wiring/rewiring dynamics is addressed as \emph{disclination dance}~\cite{Carenza22065} and is characterized by the oscillation of the angular velocity between positive and negative values (dark blue line in Fig.~\ref{fig-4}(a)), evidence of the switching behavior.

For $N \geq 4$, instead, the shape of the disclination lines is much less regular, even at very small active torques ($\bar{\zeta} \leq -0.003$), and the droplet cannot support regular motion. Active flows, sustained by the injection of angular momentum, strengthen the deformations induced by chirality  and the dynamics becomes chaotic (blue diamonds in Fig.~\ref{fig-3}(a)).
However, such regime is reached for any value of $N$ for strong enough values of the in-warding torque dipole $|\bar{\zeta}|$, and is characterized by the proliferation of new pairs of oppositely charged semi-integer point defects  on the surface, connected by disclination lines in the droplet bulk
%As chirality is increased the rotational regime is found to set up at lower values of $\bar{\zeta$}.  By increasing the active torque parameter over a critical threshold $|\bar{\zeta}|>12\times 10^{-3}$,
  (Fig.~\ref{fig-3}(d)).
Indeed, in this chaotic regime, no closed loop develops in the interior of the droplet regardless of the intensity of both $\bar{\zeta}$ and $N$. This is a fundamental difference with respect to the case of a droplet fueled by active force dipoles, where both open disclination lines (connecting surface defects) and closed loops (in the bulk) were found.

Interestingly, we find that the chaotic dynamics of the disclination lines does not lead to the motion of the droplet. Nevertheless, the surface defects undergo a non-trivial evolution, continuously nucleating and annihilating in pairs, under the effect of active forcing and elastic repulsion. This leads to a chaotic rotational dynamics, as suggested by the behavior of the angular velocity $\vec{\omega}$ for the case at $\bar{\zeta}=-0.011$ and $N=1$, plotted in Fig.~\ref{fig-4}(a).
The absence of translational motion is related to the symmetry properties of the shape attained by the droplet in the turbulent regime.  
Indeed, as $|\bar{\zeta}|$ is increased from the rotational/disclination dance regime towards the chaotic region of the phase diagram, the droplet progressively looses its spherical shape and deforms. To quantify this behavior, we introduce the spherical and cylindrical deformation parameters as $d_{min}/d_{dmax}$ and $d_{min}/d_{med}$, respectively, where $d_{max} > d_{med} > d_{min}$ denote the (time-averaged) eigenvalues of the Poinsot matrix associated to the concentration field. Fig.~\ref{fig-4}(b) shows that as the intensity of the in-warding active torque is raised, the spherical parameter rapidly decreases (regardless of $N$) and the droplet attains the shape of a prolate ellipsoid, thus preserving cylindrical symmetry, since the cylindrical parameter is roughly constant and close to $1$. Hence, while overall rotations are supported by the loss of sphericity, translational motion does not occur since the overall momentum injection is globally balanced, in accordance to preserved cylindrical symmetry.

The transition to the turbulent regime is also accompanied by a discontinuity in the (time-averaged) magnitude of the angular velocity $\vec{\omega}$ signalled by a sharp jump for any value of $N$ (highlighted by a black dot in Fig.~\ref{fig-4}(c)). Moreover, $|\vec{\omega}|$ linearly increases with $|\bar{\zeta}|$ and decreases with $N$, since at lower chirality, the pattern attained by the liquid crystal is less affected by the chiral elastic interaction and its effects are suppressed with respect to local non-equilibrium energy input provided by the active torque, leading to faster rotations.

\begin{figure*}[t]
\centering
{\includegraphics[width=1\textwidth]{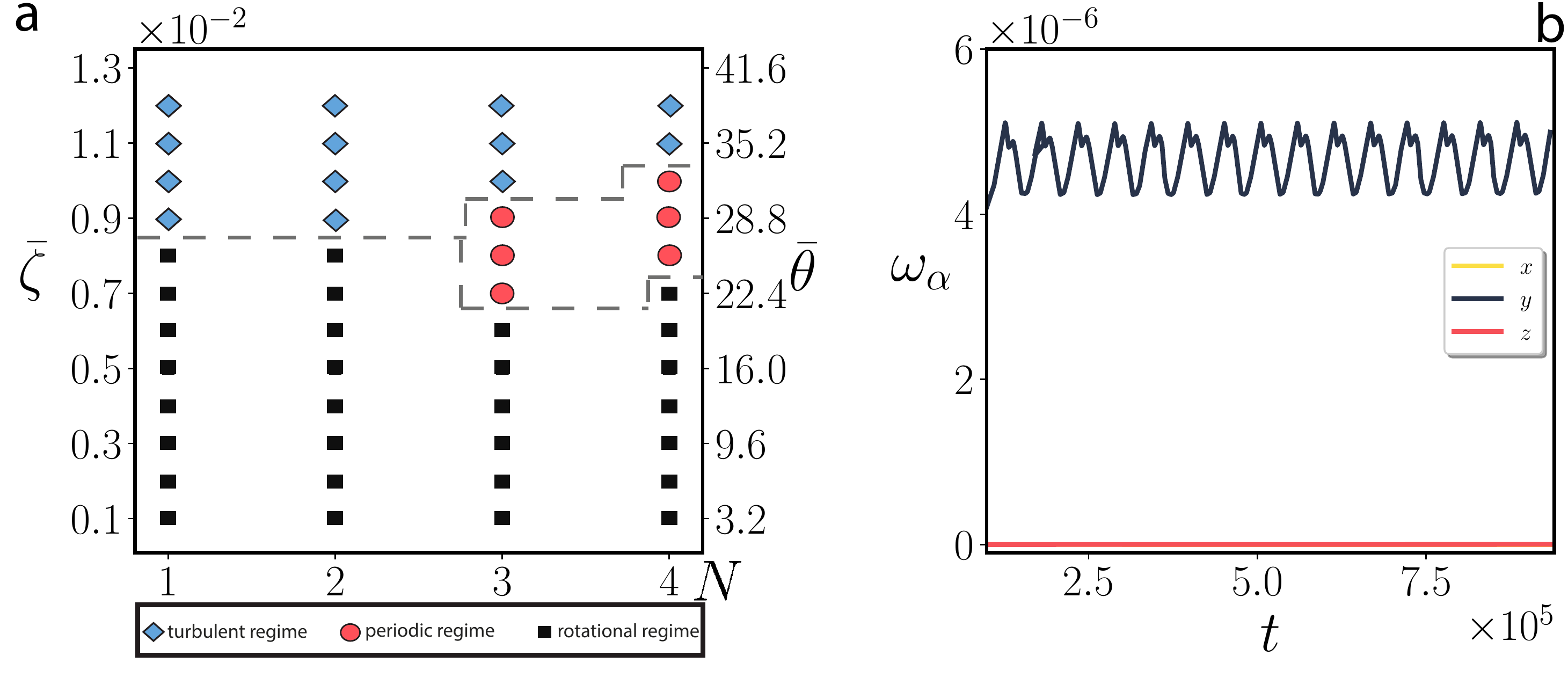}}
\caption{\textbf{Dynamical regimes of cholesteric droplets with out-warding torque dipoles ($\bar{\zeta}>0$)}: Panel (a) shows the behavior of the cholesteric droplet at varying both $N$ and $\bar{\zeta}$. Panel (b) shows the components of the angular velocity at $N=3$ and $\bar{\zeta}=7 \times 10^{-3}$ in the red region of the phase diagram.}
\label{fig-5}       % Give a unique label
\end{figure*}

\begin{figure*}[t]
\centering
{\includegraphics[width=1\textwidth]{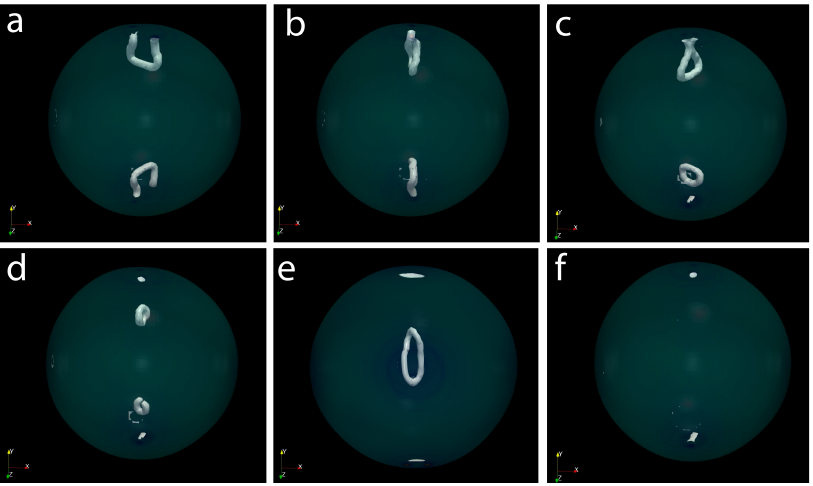}}
\caption{\textbf{Nucleation-annihilation dynamics in cholesteric droplets with out-wading torque dipoles}: Panels (a)-(f) show the evolution of the disclination lines during a cycle of the periodic nucleation-annihilation dynamics found in droplets at large $N$. First, the two antipodal boojums split to form two couples of semi-integer defects (panel (a)) which undergo rotational dynamics (panel (b)). Sustained by activity, each couple of defects rejoin to form a boojum at the poles (signaled by the white spots at the droplets poles), leaving the droplet bulk populated by ring disclinations (panel~(c). These migrate towards the center of the droplet (panel (d)), merge with each other (panel (e)) and finally annihilate (panel (f)), restoring the initial twisted bipolar configuration.}
\label{fig-6}       % Give a unique label
\end{figure*}

\subsection{Out-warding active torque dipoles}
\label{sec-6.2}
In this Section we will address the response of our cholesteric droplet to the active forcing provided by out-warding torque dipoles ($\bar{\zeta}>0$).
At small $\bar{\zeta}$ (black squares in Fig.~\ref{fig-5} (a)) a rotational regime with similar features as in the previous paragraph is found. The range of stability of such regime is extended with respect to the case with in-warding torque dipoles, since in the present case there is no competing mechanism between the handedness of the cholesteric structures developed by the LC and the flow induced by non-equilibrium flows.
Interestingly, the transition towards the chaotic state (blue diamonds in Fig.~\ref{fig-5} (a)) occurs at lower $\bar{\zeta}$ for LC with weaker chirality.

While for $N \leq 2$ we observe a direct transition from the rotational to the turbulent state at $\bar{\zeta}=9\times 10^{-3}$, a new dynamical regime appears if LC with shorter pitch are considered.
Indeed, for $N \geq 3$ and strong enough active torques (see red circles in Fig.~\ref{fig-5} (a)) the droplet enters a state characterized by a periodic dynamics, summarized in panels~(a)-(f) of Fig.~\ref{fig-6} (and supplementary movie).
Starting from the twisted bipolar configuration--characteristic of the rotational regime--each  boojum located at the droplet poles, splits into two defects of $+1/2$ charge (panel~(a)), giving rise to a short disclination line. First, the semi-integer defects slightly move apart and undergo a precessional dynamics around each other (panels (a)-(b)), then, under the effects of the attraction due to the active forcing, the two surface defects tend to rejoin into a $+1$ defect at each pole (white spots in panels (c)), resulting in the formation of two small disclination loops in the bulk (panel (c)). These migrate towards the center of the droplet (panel (d)) with two boojums at the poles, merge with each other (panel (e)) and finally annihilate (panel (f)), leaving the droplet in the initial twisted bipolar state.

Such intermittent dynamics exhibits relevant differences from the disclination dance regime found at $\bar{\zeta}<0$. First, in the present case this periodical state only appears at large enough chirality and second, the direction of rotation is preserved during the evolution, as shown by panel~(b) of Fig.~\ref{fig-5}, where the components of the angular velocity are plotted for the case at $N=3$ and $\bar{\zeta}=7 \times 10^{-3}$.

\section{Conclusions}
\label{sec-7}

In this paper we presented an analysis of the chaotic dynamics of an active cholesteric droplet fueled by active force and torque dipoles, in which we varied both the intensity of the activity parameter and cholesteric power.
In the case of a droplet fueled by force dipoles only, we found that the onset of turbulence is driven by the splitting of antipodal boojums into $+1/2$ point defects, paired by disclination lines that pierce the interior of the droplet. By increasing the intensity of the cholesteric power we find that disclination loops may nucleate (and eventually annihilate) in the bulk of the droplet. Their chaotic dynamics leads to the symmetry breaking of the LC arrangement, which is the fundamental ingredient to the development of the erratic motion of the droplet. 

Cholesteric droplets fueled by in-warding active torques exhibit three different dynamical regimes which can be selected by tuning both thermodynamic chirality and the intensity of angular momentum injection. At small $|\bar{\zeta}|$ the droplet develops a stable rotational motion  around the axis defined by two boojums at the poles. For small chirality ($N \leq 3$), a disclination dance regime is found at intermediate values of $|\bar{\zeta}|$, characterized by the wiring/rewiring dynamics of two disclination lines. By further increasing the intensity of the in-warding torque dipoles the droplet undergoes a transition from the rotational towards a turbulent state with the proliferation of semi-integer defects on the surface of the droplet connected by line disclinations which pierce the droplet bulk, leading to a chaotic rotational dynamics.
We found that higher chirality ($N \geq 4$) determines the suppression of the intermittent disclination dance dynamics and the droplet directly undergoes the transition to the chaotic regime from the rotational state.

Switching the sign of the active torque $\bar{\zeta}$, we found that out-warding torque dipoles favor the transition to the turbulent state at weaker chirality ($N \leq 2$), while at higher chirality a periodical dynamics characterized by the nucleation/annihilation of pairs of ring disclination set up at intermediate active forcing, before entering the chaotic state at higher $\bar{\zeta}$.

\section{Acknowledgements}
The work has been performed under the Project HPC-EUROPA3 (INFRAIA-2016-1-730897), with the support of the EC Research Innovation Action under the H2020 Programme; in particular, the authors gratefully acknowledges the support and the computer resources  provided by EPCC.

%For one-column wide figures use syntax of figure~\ref{fig-1}
%\begin{figure}[h]
%\centering
%\includegraphics[width=1cm,clip]{tiger}
%\label{fig-1}       % Give a unique label
%\end{figure}

%For two-column wide figures use syntax of figure~\ref{fig-2}
%\begin{figure*}
%\centering
%\vspace*{5cm}       % Give the correct figure height in cm
%\label{fig-2}       % Give a unique label
%\end{figure*}

%For figure with sidecaption legend use syntax of figure
%\begin{figure}
%\centering
%\sidecaption
%\includegraphics[width=5cm,clip]{tiger}
%\label{fig-3}       % Give a unique label
%\end{figure}

%For tables use syntax in table~\ref{tab-1}.

%
% BibTeX or Biber users please use (the style is already called in the class, ensure that the "woc.bst" style is in your local directory)
\bibliographystyle{unsrt}
\bibliography{refs}
%\begin{thebibliography}{}
%%
%% and use \bibitem to create references.
%Journal Author, Journal \textbf{Volume}, page numbers (year)
%Book Author, \textit{Book title} (Publisher, place, year) page numbers
%\end{thebibliography}

\end{document}